\documentclass[a4paper, 10pt, conference]{IEEEtran}

\usepackage{graphicx}
\usepackage{cite}
\usepackage{amsmath,amssymb,amsfonts}
\usepackage{algorithmic}
\usepackage{textcomp}
\usepackage{lipsum}
\usepackage{url}
\usepackage{array}
\usepackage{booktabs}
\usepackage{multirow}
\usepackage[table]{xcolor}
\usepackage{colortbl}
\usepackage{adjustbox}
\usepackage{enumitem}

\usepackage{subcaption} 
\usepackage{caption} 

\definecolor{HighPriority}{HTML}{FFC7CE}   
\definecolor{MediumPriority}{HTML}{FFEB9C} 
\definecolor{LowPriority}{HTML}{C6EFCE}    

\begin{document}

\title{Driving Privacy Forward: Mitigating Information Leakage within Smart Vehicles through Synthetic Data Generation}

\author{
\IEEEauthorblockN{Krish Parikh}
\IEEEauthorblockA{krish.parikh@warwick.ac.uk\\ \today\\}
}

\maketitle

\begin{abstract}
Smart vehicles produce large amounts of data, much of which is sensitive and at risk of privacy breaches. As attackers increasingly exploit anonymised metadata within these datasets to profile drivers, it's important to find solutions that mitigate this information leakage without hindering innovation and ongoing research. Synthetic data has emerged as a promising tool to address these privacy concerns, as it allows for the replication of real-world data relationships while minimising the risk of revealing sensitive information. In this paper, we examine the use of synthetic data to tackle these challenges. We start by proposing a comprehensive taxonomy of 14 in-vehicle sensors, identifying potential attacks and categorising their vulnerability. We then focus on the most vulnerable signals, using the Passive Vehicular Sensor (PVS) \cite{9277846} dataset to generate synthetic data with a Tabular Variational Autoencoder (TVAE) \cite{tvae} model, which included over 1 million data points. Finally, we evaluate this against 3 core metrics: fidelity, utility, and privacy. Our results show that we achieved 90.1\% statistical similarity and 78\% classification accuracy when tested on its original intent, while also preventing the profiling of the driver.
\end{abstract}

\section{Introduction}
\label{sec:introduction}

\subsection{What is Synthetic Data?}

Synthetic data is artificially generated information that mimics the statistical properties and patterns of real-world data without revealing sensitive details from the original dataset. It is created using algorithms and models trained on actual data, serving as a proxy that maintains the utility of real data while protecting privacy and confidentiality.

\begin{figure}[h]
  \centering
  \includegraphics[height=6.2cm]{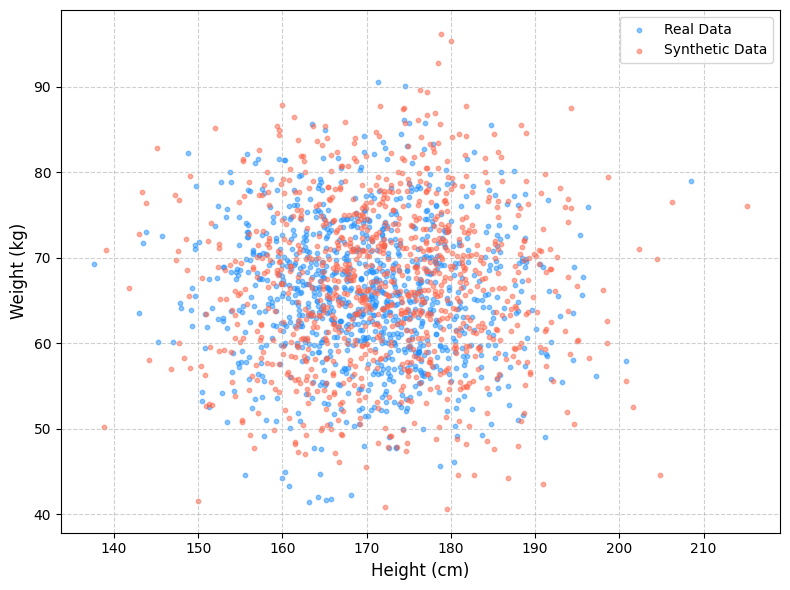}
  \caption{The synthetic data retains the structure of the original data but is not the same}
  \label{realvssynthetic}
\end{figure}

As shown in Figure \ref{realvssynthetic} the generation process involves understanding the underlying distributions, correlations, and relationships within the real data and producing new data points that reflect these characteristics. This approach is valuable when access to real data is limited due to privacy concerns, regulatory restrictions, or the potential for information leakage.

\subsection{Brief History of Synthetic Data}

The concept of synthetic data has evolved over several decades, intersecting fields such as statistics, computer science, and artificial intelligence. Key milestones include:

\begin{itemize}
    \item \textbf{Early 1990s}: Initial ideas emerged from the need to protect individual privacy in statistical databases. Researchers explored methods to release useful data without compromising personal information.
    \item \textbf{1993}: Donald B. Rubin introduced the concept of multiple imputation \cite{rubin1987} for missing data, where missing values are predicted several times by generating multiple plausible values for each missing data point. These values are drawn from a distribution based on the observed data, and the variation between them accounts for the uncertainty inherent in missing data.
    \item \textbf{Late 1990s to Early 2000s}: Government agencies, such as the U.S. Census Bureau, began experimenting with synthetic data to publish public-use microdata files, aiming to provide researchers with valuable data while safeguarding respondent confidentiality \cite{Abowd2002DisclosureLI}.
    \item \textbf{2010}: The first publicly available synthetic dataset applied to autonomous driving tasks was the Foggy Road Image Database (FRIDA) generated using the Sivic software \cite{tarel2010improved}. This dataset provided synthetic images of road scenes under varying fog conditions, enabling researchers to develop and test algorithms for visibility enhancement and object detection in adverse weather.
    \item \textbf{2014}: Ian Goodfellow and colleagues introduced Generative Adversarial Networks (GANs), enabling the creation of more realistic and complex synthetic data by having two neural networks—the generator and the discriminator—compete against each other \cite{goodfellow2014generative}.
    \item \textbf{Mid-2010s to Present}: Advances in machine learning and models like Variational Autoencoders (VAEs) revolutionized synthetic data generation across domains such as images, text, and time-series data \cite{kingma2014auto}. With increasing concerns about data privacy and regulations like the General Data Protection Regulation (GDPR), synthetic data gained prominence as a solution to share and analyze data without violating privacy laws.
\end{itemize}

\subsection{What Synthetic Data Isn't}

Despite its potential benefits, synthetic data is often misunderstood. Several misconceptions need clarification, and real-world incidents highlight the consequences of these misunderstandings.

\subsubsection{\textbf{Not Inherently Private}}

There is a common misconception that synthetic data is automatically private and safe from the risks associated with real data. However, if the process of generating synthetic data is not carefully designed, it can still expose sensitive information from the original dataset.

For example, synthetic data can be vulnerable to membership inference attacks, where an attacker can determine whether a specific data point was part of the training data \cite{hayes2019logan}. This risk exists because the synthetic data may unintentionally reveal patterns specific to individuals in the original dataset.

\subsubsection{\textbf{Not a Complete Replacement for Real Data}}

Synthetic data approximates real data but may not capture all its complexities and nuances, potentially leading to models that perform poorly when deployed in real-world settings.

In the healthcare sector, models trained exclusively on synthetic medical data may not perform adequately when applied to real patient data. A study found that while synthetic electronic health records could be useful for preliminary research, models trained solely on them required fine-tuning with real data to achieve acceptable performance levels \cite{baowaly2019}. This gap is due to synthetic data lacking subtle patterns and anomalies present in actual patient records.

\subsubsection{\textbf{Does Not Eliminate All Risks}}

Using synthetic data does not absolve organizations from adhering to data protection regulations and ethical considerations. The generation process itself can introduce risks if not properly managed.

For example, if synthetic data is generated without adequate privacy-preserving techniques, it may still be subject to re-identification attacks. Organizations may face legal challenges if the synthetic data violates regulations like the GDPR.

\subsubsection{\textbf{Not Universally Applicable}}

Synthetic data is more suitable for some applications than others. In domains where data is highly unstructured or complex, generating high-quality synthetic data remains a challenge.

In natural language processing, generating synthetic text data that accurately reflects the diversity and context of human language is difficult. Language models trained on sensitive data can inadvertently reproduce personal information. Studies have shown that large language models can regurgitate chunks of training data, including sensitive or private information, posing significant privacy risks \cite{274574}.

\subsection{Conclusion}

Understanding what synthetic data is—and what it isn't—is crucial for its effective and responsible use. While it offers significant advantages, synthetic data is not a cure-all for data privacy and utility challenges. Careful consideration of the generation process, potential risks, and appropriate use cases is essential to prevent real-world mishaps and harness the true potential of synthetic data.

\section{Why Does It Matter?}

The significance of synthetic data becomes apparent when considering the limitations of traditional data anonymisation techniques and the increasing risks of privacy breaches in the era of big data. This section illustrates these challenges through real-world examples and emphasises the necessity for more robust solutions like synthetic data.

\subsection{Analogy: The Story of an Unfortunate Patient}

Let's consider the case of Bob, a participant in a medical census conducted by a hospital. Believing his personal information to be secure after his name and direct identifiers were removed, Bob assumed his privacy was protected. However, a curious researcher was able to use auxiliary information, such as age, sex, and zip code to re-identify Bob and confirm his medical records. By leveraging quasi-identifiers—attributes like age, gender, and ZIP code that can be linked to external data sources such as voter registration information—the researcher successfully inferred sensitive details about Bob's health conditions, which were passed on to his insurance company and increased their insurance premium the following month.

While this scenario, is hypothetical, the potential to re-identify Bob is not. Latanya Sweeney's seminal work demonstrated that 87\% of the U.S. population could be uniquely identified using just three pieces of information: ZIP code, birth date, and gender \cite{sweeney2018}. Such re-identification attacks highlight the vulnerabilities in traditional anonymisation methods.

\subsection{The Problem with Anonymisation}

Traditional methods of data anonymisation, such as data masking, generalisation, and suppression, often do not provide sufficient privacy protection. Techniques like \(k\)-anonymity aim to make each record indistinguishable from at least \(k-1\) other records by generalising or suppressing certain attributes \cite{Samarati1998ProtectingPW}. However, these methods have limitations, mainly a loss of data utility.

Below are two tables: Table \ref{tab:original_dataset} containing personally identifiable information and Table \ref{tab:kanonymity_dataset} having applying \(k\)-anonymity.

\begin{table}[h!]
\centering
\begin{tabular}{|l|c|c|c|}
\hline
\multicolumn{4}{|c|}{\textbf{Original Dataset}} \\ \hline
\textbf{Age} & \textbf{Sex} & \textbf{ZIP} & \textbf{Disease} \\ \hline
28 & F & 23467 & Cancer \\ \hline
17 & M & 12345 & Heart Disease \\ \hline
34 & M & 65490 & Flu \\ \hline
41 & M & 84933 & Bronchitis \\ \hline
\end{tabular}
\caption{Original Dataset}
\label{tab:original_dataset}
\end{table}

\begin{table}[h!]
\centering
\begin{tabular}{|l|c|c|c|}
\hline
\multicolumn{4}{|c|}{\textbf{\(k\)-Anonymised Dataset}} \\ \hline
\textbf{Age} & \textbf{Sex} & \textbf{ZIP} & \textbf{Disease} \\ \hline
10--29 & * & * & Cancer \\ \hline
10--29 & * & * & Heart Disease \\ \hline
30--49 & M & * & Flu \\ \hline
30--49 & M & * & Bronchitis \\ \hline
\end{tabular}
\caption{\(k\)-Anonymized Dataset}
\label{tab:kanonymity_dataset}
\end{table}

Over-generalization and suppression can make data less useful for analysis. For example:

\begin{itemize}
    \item \textbf{Age Generalisation}: Precise age values are replaced with broad age ranges (e.g., ``28'' becomes ``10--29''), which may obscure age-related trends and patterns critical for certain studies.
    \item \textbf{Attribute Suppression}: Specific attributes like ``Sex'' and ``ZIP'' are suppressed (represented by ``*''), eliminating the ability to analyse data based on gender or geographic location.
\end{itemize}

This loss of detailed information hampers the ability to identify meaningful correlations and can significantly reduce the dataset's utility for research and analysis.

Moreover, \(k\)-anonymity does not fully protect against re-identification attacks, especially when attackers have access to auxiliary information that can be cross-referenced with the anonymised dataset \cite{machanavajjhala2007diversity}. For instance:

\begin{itemize}
    \item \textbf{Background Knowledge Attack}: If an attacker knows that a 28-year-old female from ZIP code 23467 was in the dataset, they might still link the record to the individual, even with generalized data.
    \item \textbf{Homogeneity Attack}: When all records in an anonymised group share the same sensitive attribute value (e.g., "Disease"), the specific information can be inferred despite the generalisation.
\end{itemize}

\subsection{The Opportunity for Synthetic Data}
Given the shortcomings of traditional anonymisation, synthetic data presents a promising alternative. By generating data that captures the statistical properties of the original dataset without containing personally identifiable records, synthetic data can:
\begin{itemize} 
    \item \textbf{Enhance Privacy Protection}: Synthetic data minimises the risk of re-identification by excluding real individuals' records, thus reducing the likelihood of linkage attacks with external datasets. 
    \item \textbf{Maintain Fidelity}: Generators can learn the underlying relationships and patterns in the original data, ensuring statistical similarity.
    \item \textbf{Improve Model Performace}: By generating new samples of synthetic data, this can effectively be coupled with real world data to increase the size of the training dataset, enabling better model performance.
\end{itemize}

In the context of Bob's story, had the hospital provided researchers with synthetic data instead of anonymised real data, the risk of re-identification would have been significantly reduced. Moreover, as data privacy regulations like the GDPR impose stricter requirements on personal data processing, synthetic data offers a compliant way to leverage data for analytics and AI development.

Now that we understand what synthetic data is and its role in privacy, we can move on to the issue at hand, information leakage within smart vehicles.

\section{The Problem at Hand}

The rapid advancement of smart vehicle technology has led to a significant increase in the number of connected cars on the road. As of 2023, recent estimates indicate that there are over 237 million connected vehicles worldwide, a figure projected to grow to over 850 million by 2035 \cite{statista2023}. These vehicles collect and transmit vast amounts of data related to driver behaviour, vehicle performance, and environmental conditions. While this data is instrumental in enhancing vehicle functionality, safety features, and user experience, unauthorised access to this data raises serious privacy concerns.

\subsection{How Secure is Your Data?}

Although one might assume that in-vehicle data remains secure within the vehicle, a study conducted by the Mozzila Foundation shows otherwise. It was found that 84\% of car brands researched share personal data with service providers, data brokers, and other unknown entities, while 76\% admit to selling this personal data \cite{caltrider2023}.

\begin{figure}[h]
  \centering
  \includegraphics[height=4.5cm]{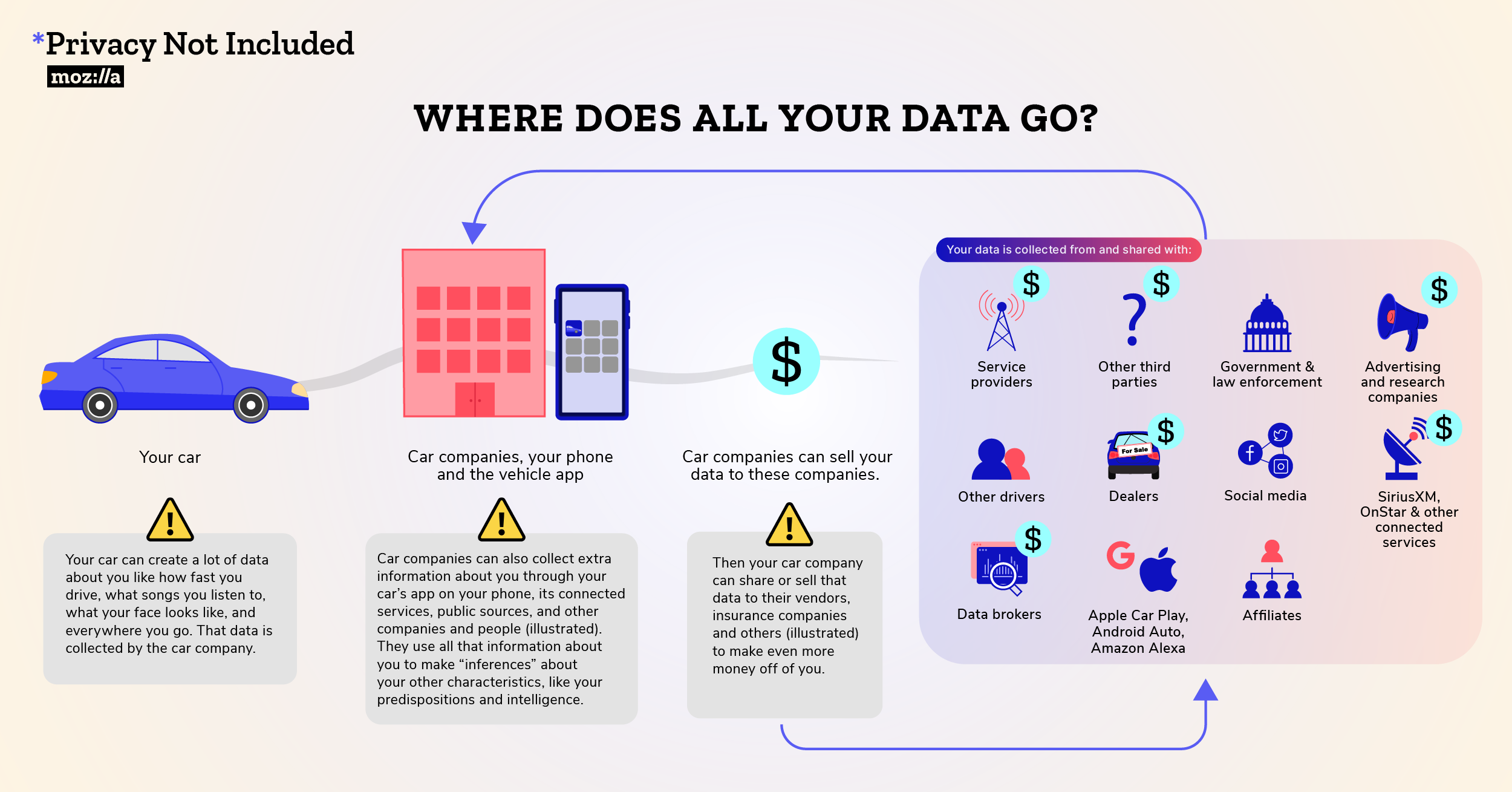}
  \caption{Data flows from within your vehicle system to various entities such as manufacturers, service providers, and third-parties \cite{caltrider2023}.}
  \label{data_flow}
  
\end{figure}

Even when manufacturers do not sell your data, significant data leaks have occurred. One example is Toyota Motor Corporation, which experienced a decade-long data breach in its cloud environment, exposing the location information of 2.15 million customers \cite{toyota_leak}.

As shown in Figure \ref{data_flow}, numerous organisations rely on the personal data collected by your car. This then raises the question: what makes this data so valuable?

\subsection{Why Your Data is Valuable?
}

While the specific intent of data collection on smart vehicles can vary depending on the type of information being targeted, in general, attackers and organisations aim to gain additional knowledge about the driver of the vehicle. This knowledge can range from collecting explicit data points, such as location data to determine the driver’s driving patterns, to inferring derived data points, such as determining a driver’s political leanings from characteristics like their engine size \cite{Guess_Which_Car_Type_I_Am_Driving}.

When such sensitive data is intentionally or unintentionally released into an untrusted environment, it leads to information leakage. Given the virtually limitless scope of derived data, due to the numerous statistical relationships that in-vehicle data points have with external data, this raises concerns about unintended information leakage. During the distribution of this data, the potential vulnerabilities and sensitive information that could be inferred are often unclear. Hence, despite being anonymised and deemed private, such methods are not sufficient and can still lead to driver profiling.

\subsubsection{\textbf{Driver Profiling - The Bad}}

One such vulnerability was highlighted by researchers at the University of Padova. They demonstrated that by observing the state of charge of an electric vehicle during the charging phase. They could use this exchange of the electric current between the vehicle and EVSE to uniquely identify the vehicle. A classification model was then trained to identify where the vehicle charged, achieving 0.9 recall and 0.85 precision, and thus inferring the drivers driving patterns \cite{Tell_Me_How_You_Re-Charge}. This knowledge could then go on to be used by advertising agencies to perform targeted advertising to the driver, given that they know the locations in which the driver charges.

\subsubsection{\textbf{Driver Profiling - The Good}}

It seems inevitable that driver profiling will become more common, given the large amount of data that can be inferred about an individual. However, it’s important to note that this practice is not always negative. One example of a positive application is in anti-theft technology. Research has shown that drivers have unique driving patterns that can be characterised by attributes such as speed and steering behaviour \cite{know_your_master}. A classification model could be trained to recognise these patterns and detect when an unauthorised person is driving the vehicle. If a thief were to steal the car, the model could quickly immobilise the vehicle as it would fail to recognise the unauthorised driver.

\subsubsection{\textbf{Finding The Middle Ground}}

Given both the risks and potential innovation that come from driver profiling. We need to be able to determine what in-vehicle sensors are most vulnerable to information leakage and the significance of an attack in which this data is then used by an adversary. In such cases, we will seek to use synthetic data to address this vulnerability. In the following section, I have proposed a comprehensive taxonomy outlining the key sensors within the vehicle and their potential to leak information.

\section{Information Leakage Taxonomy}
\label{sec:taxonomy}

It is important to understand which sensors in smart vehicles are vulnerable and to assess the extent of potential information leakage before generating synthetic data to mitigate the risks of driver profiling. As shown in Table \ref{tab:taxonomy}, a total of 14 in-vehicle sensor signals have been identified as having the potential to leak sensitive information. These signals include obvious ones like GPS data and camera feeds, as well as less obvious sources such as accelerometer readings and battery management data.

\begin{table*}[htbp]
\caption{Smart Vehicle Signal Taxonomy Grouped by Priority}
\label{tab:taxonomy}
\centering
\begin{adjustbox}{max width=\textwidth}
\normalsize 
\begin{tabular}{|p{4.5cm}|p{5cm}|p{7cm}|p{3.5cm}|}
\hline
\rowcolor{gray!20}
\textbf{Signal (Source)} & \textbf{Type} & \textbf{Potential Information Leakage} & \textbf{Category} \\
\hline

\rowcolor{HighPriority}
GPS Data (GPS Module) & Longitude and latitude coordinates, speed, heading (direction) & Live location tracking, habitual routes, travel patterns and habits & Vehicle Telematics Signal \\
\hline

\rowcolor{HighPriority}
Camera Data (Vehicle Cameras) & Video streams, image captures & Visual information of vehicle environment, captures passengers, pedestrians, license plates, location identification from images & Environment and External Interaction Signal \\
\hline

\rowcolor{HighPriority}
Bluetooth/Wi-Fi Data (Communication Modules) & Device connections, data transfers & Device pairing information (MAC addresses) uniquely identifying devices, data transfer logs exposing personal files or communications, determining driver location via Bluetooth scanning & Connectivity and Communication Signal \\
\hline

\rowcolor{HighPriority}
Cellular Data (Cellular Modem) & Communication logs, data usage & Communication logs, remote diagnostics access, potential unauthorized access to vehicle systems & Connectivity and Communication Signal \\
\hline

\rowcolor{MediumPriority}
Engine Control Unit Data (ECU) & Engine performance, fuel consumption, RPM, safety systems, energy management & Aggressive acceleration patterns, braking behavior, overall driving style & Vehicle Telematics Signal \\
\hline

\rowcolor{MediumPriority}
Control Area Network Bus Data (CAN Bus) & Messages between vehicle subsystems including: Sensor data, telematics data, powertrain data, infotainment system data, comfort and convenience system data & Vehicle operational data, driver behavior, vehicle status, passenger information (e.g., seatbelt usage) & In-Vehicle Network Signal \\
\hline

\rowcolor{MediumPriority}
On-Board Diagnostic Data (OBD Port) & Diagnostic trouble codes, emissions data, engine performance, fuel efficiency, ignition system data, transmission data, sensor data, battery data & Reveals vehicle issues and maintenance history, sensor readings inferring performance and driver behavior & In-Vehicle Network Signal \\
\hline

\rowcolor{MediumPriority}
Infotainment System Data (Infotainment Unit) & Media playback history, navigation usage, connectivity status & Media preferences, travel destinations, paired devices and communication logs & User Interaction and Infotainment Signal \\
\hline

\rowcolor{MediumPriority}
Human-Machine Interface Data (Controls) & User inputs, voice commands & Voice recordings capturing sensitive information, user interaction patterns & User Interaction and Infotainment Signal \\
\hline

\rowcolor{MediumPriority}
Accelerometer/Gyroscope Data (IMU Sensors) & Orientation, acceleration & Detailed driving behavior (e.g., driver aggression), potential to uniquely identify driver & Sensor Data \\
\hline

\rowcolor{MediumPriority}
Battery Management System Data (BMS) & Battery charge status, voltage, temperature & Charging patterns revealing user habits, battery status and health information & Power and Energy Management Signals \\
\hline

\rowcolor{LowPriority}
Lidar/Radar Data (Sensors) & Point cloud sensor readings & Environmental mapping around vehicle, object detection, potential inference of surroundings & Environment and External Interaction Signal \\
\hline

\rowcolor{LowPriority}
Temperature Data (Environmental Sensors) & Ambient conditions & Infer vehicle's location or route based on environmental conditions, users' or passengers' comfort preferences & Sensor Data \\
\hline

\rowcolor{LowPriority}
Energy Consumption Data (Vehicle Systems) & Power usage by components & Power usage patterns, vehicle usage and efficiency, detailed breakdown of energy consumption & Power and Energy Management Signals \\
\hline
\end{tabular}
\end{adjustbox}

\vspace{1em}
\noindent\begin{tabular}{l l}
\rowcolor{HighPriority} \hspace{1em} & \textbf{High Priority (Red):} Directly identifies individuals or tracks their behavior. \\
\rowcolor{MediumPriority} \hspace{1em} & \textbf{Medium Priority (Amber):} Infers behaviors but doesn't directly identify individuals. \\
\rowcolor{LowPriority} \hspace{1em} & \textbf{Low Priority (Green):} Provides general information without identifying individuals. \\
\end{tabular}

\end{table*}

The taxonomy prioritises vehicle sensors based on three criteria:

\begin{enumerate} 
    \item \textbf{Identifiability of Driver Behavior}: Data that directly identifies individuals or tracks their behaviour is considered more vulnerable than data that requires further inference, such as being coupled with auxiliary data to perform driver profiling.
 
    \item \textbf{Obtainability of Data}: Data deemed readily accessible and obtainable by an attacker is considered more vulnerable. For example, despite infotainment data being able to identify the driver, much of this interaction is encrypted, making access to this data relatively difficult for an attacker.
    
    \item \textbf{Intent of Profiling}: While intentions vary and can never truly be predicted, we identified the most significant threats to driver safety based on known attacks. For instance, while power usage patterns could be used by energy suppliers to understand consumption habits, they were not deemed as vulnerable as GPS data, which could be used by an attacker for stalking.
    
\end{enumerate}

Surprisingly, obtaining access to in-vehicle data is not just limited to manufacturers, data purchases, or illegal activities. There are completely legal and low-cost methods that enable driver profiling for anyone. For instance, studies have demonstrated that Bluetooth Low Energy (BLE) scanners positioned at different traffic lights can intercept the MAC addresses of passing vehicles. By collecting these unique identifiers at multiple locations, it becomes possible to track a specific vehicle's movements across a city. This method is particularly concerning because it does not require illegal access to the vehicle's systems or cooperation from manufacturers—it's a legal and relatively straightforward way to identify and monitor car locations \cite{Bluetooth-Based}.

Unauthorised access to a driver's location has serious implications of:

\begin{itemize}
    \item \textbf{Stalking and Harassment}: Individuals could be targeted and followed without their knowledge.
    \item \textbf{Home Invasion and Burglary}: Knowing when a person is away from their home can aid in planning criminal activities.
    \item \textbf{Targeted Advertising}: Visits to frequent locations and understanding driver behaviour could be used for targeted advertising.
\end{itemize}

Given this direct identification of driver behaviour, obtainability of data and malicious intent of profiling, it became evident that protecting location data is paramount. Therefore, in the process of synthetic data generation, I concentrated on creating realistic yet privacy-preserving GPS data. By generating synthetic location data, the goal is to preserve the functionality of vehicle systems that rely on such information while safeguarding individual privacy and minimising the risks associated with data leakage.

\section{Synthetic Data Generation} 

\begin{figure}
    \centering
    \includegraphics[width=1.05\linewidth]{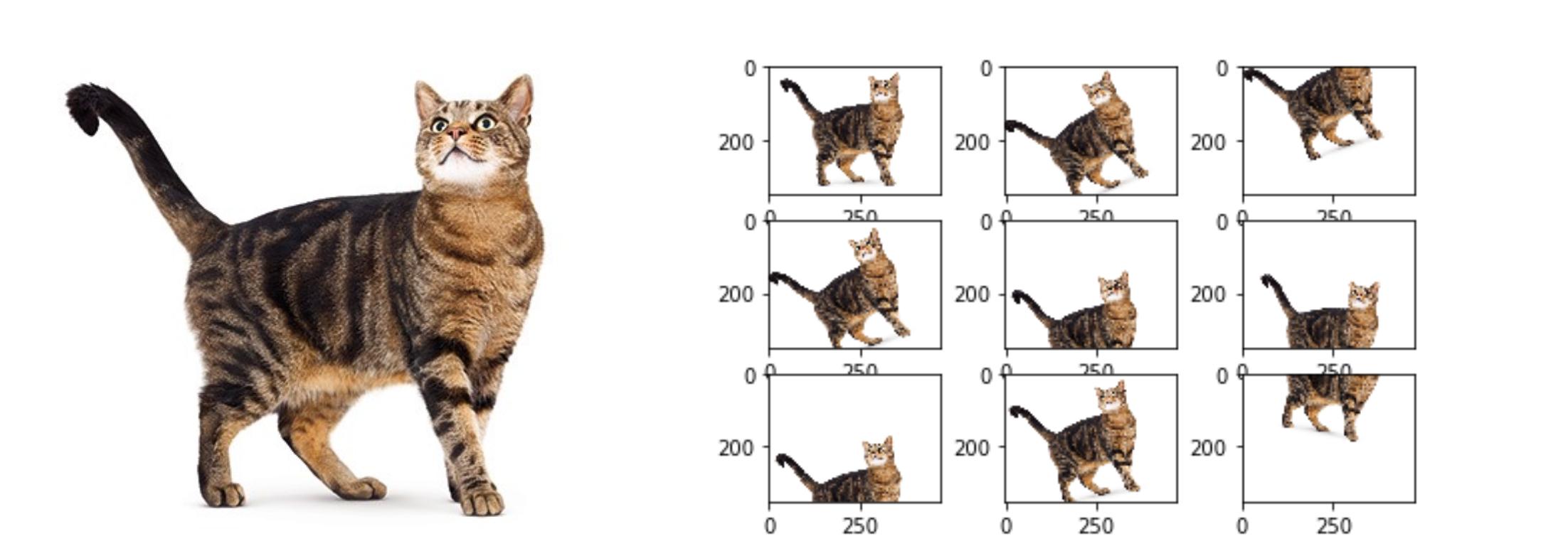}
    \caption{Typical data augmentation example using rotation, reflection, and translation of an image \cite{ubiai2023}.}
    \label{fig:augmentation}
\end{figure}

Having understood the benefits of synthetic data and the challenges at hand, we now turn to finding an effective solution. The question becomes: where do we start? Synthetic data generation is a broad concept encompassing a wide range of techniques for creating artificial data. The generation process can vary from using Instruct LLMs to create entire datasets from scratch to simple augmentation techniques like rotation and inversion of images, as illustrated in Figure \ref{fig:augmentation}.

In general, synthetic data generators can be categorised into the following techniques:

\begin{itemize}
    \item \textbf{Generative AI} - Data is generated using deep learning techniques, such as Generative Pre-trained Transformers (GPT), Generative Adversarial Networks (GANs), or Variational Auto-Encoders (VAEs), which learn the underlying distribution of real data to generate similarly distributed synthetic data.
    \item \textbf{Rules Engine} - Artificial data is created based on user-defined business rules. Intelligence can be added by referencing relationships between data elements to maintain relational integrity. 
    \item \textbf{Data Masking} - Personally identifiable information (PII), such as zip codes or names, is replaced with fabricated but structurally consistent values. This ensures that sensitive data cannot be linked to individuals while retaining the overall relationships and statistical characteristics of the data. 
    \item \textbf{Data Augmentation} - Techniques like flipping, rotation, scaling, and translation are applied to existing data to create new data points. This is primarily used in synthetic image generation.
    \item \textbf{Noise Injection} - Randomly sampled data points from known distributions, along with noise, are added to existing data to create new data points that closely resemble real-world data while preserving privacy.
    \item \textbf{Simulation Software} - Custom software is used to create and run simulations that generate synthetic data. For example, the video game Grand Theft Auto V has been used to acquire data for object detection of UAVs \cite{9956710}.
\end{itemize} 

While Generative AI was the first approach that came to mind for generating synthetic data, it is only a subset of the broader range of techniques available.

\subsection{Method Selection}

The method of generation one chooses depends mainly on the type of data and its intended use case. In our case, however, we are working with tabular data. Specifically, the Passive Vehicular Sensors (PVS) Dataset that contains GPS, accelerometer, gyroscope and magnetometer data collected by a \textit{Volkswagen Saveiro} driving for \textit{13.81km} on various road conditions \cite{9277846}.

\begin{figure*}[t]
    \centering
    \includegraphics[width=\textwidth, height=\textheight, keepaspectratio]{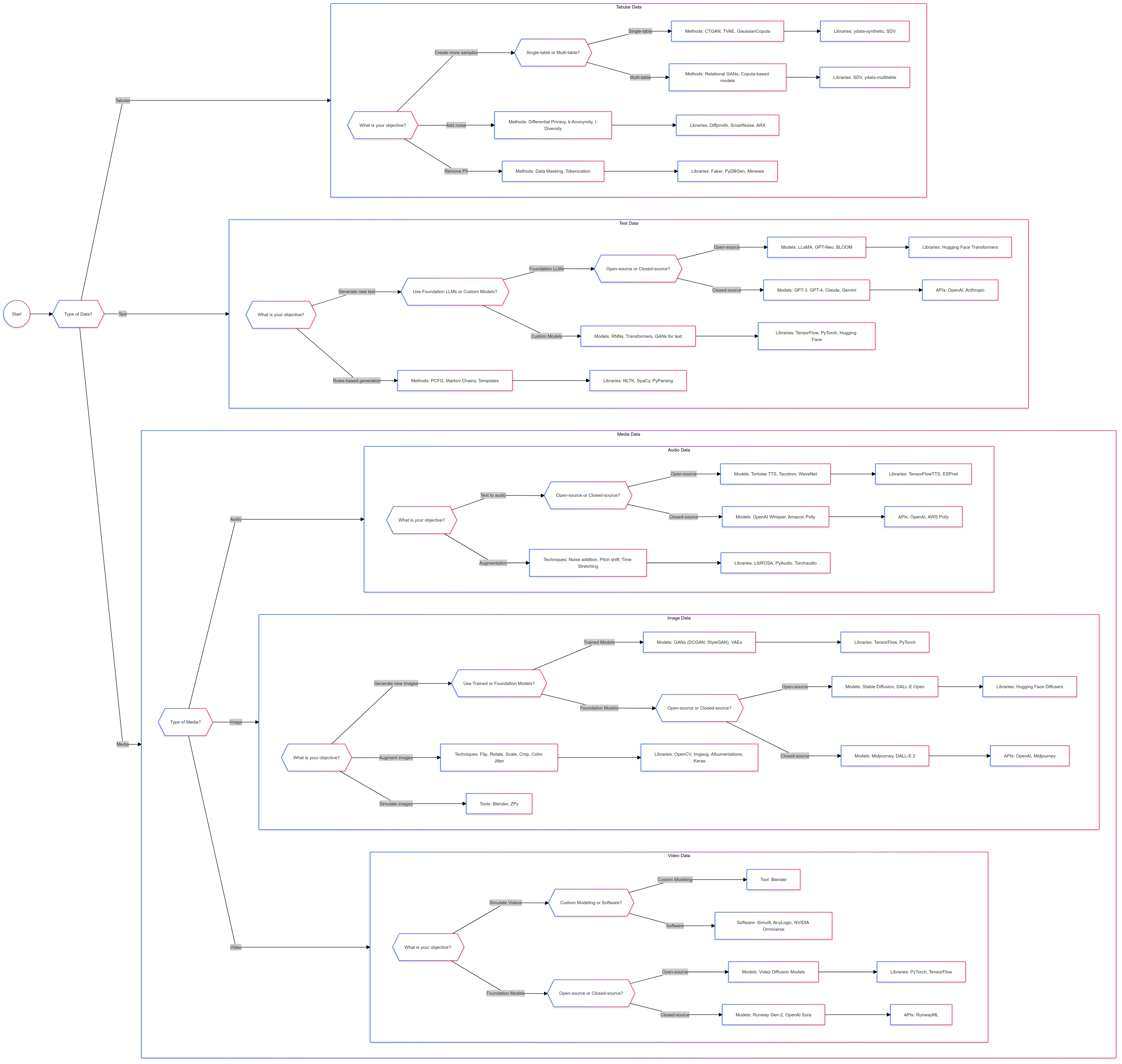}
    \caption{Describes the available methods, models and use cases for generating synthetic data}
    \label{fig:flow-chart}
\end{figure*}

Given that we are working with single table data, with the focus on being able to capture the complex relationships between data points, using the proposed flow chart in Figure \ref{fig:flow-chart}, it's clear that deep learning methods, in particular models such as GAN's and VAE’s, are most suitable for us. 

\subsection{Model Selection}

As Figure \ref{fig:flow-chart} describes, even once we have decided what method to use, we can use various models to generate our synthetic samples. We can simplify this by understanding that many of these models are flavours of both the GAN and the VAE architecture.

\subsubsection{\textbf{Variational Autoencoder (VAE)}}

A VAE is a generative model designed to learn meaningful patterns from data and use those patterns to generate new, similar data. It first compresses the input data into a simpler, lower-dimensional representation through an encoder network. This compressed representation isn't a fixed point but a distribution, meaning the model captures a range of possibilities rather than just one outcome.

Using a decoder network, the VAE decodes this compressed information back into its original form (or as close as possible). The learning process involves balancing two key objectives:

\begin{itemize}
    \item \textbf{Reconstruction Loss} - This checks how well the decoder can rebuild the original data.
    \item \textbf{KL Divergence} - This regularises the latent space by encouraging the encoded distributions to remain close to a standard normal distribution, which helps the model generalise well to new data.
\end{itemize}

\subsubsection{\textbf{Generative Adversarial Network (GAN)}}

A GAN takes a different approach to generating synthetic data by setting up a playful rivalry between two neural networks:

\begin{itemize}
    \item\textbf{Generator} - Creates synthetic data by starting with random noise and gradually learning to transform that noise into data that mimics the real-world data it has seen.
    \item \textbf{Discriminator} - Acts as a referee, learning to distinguish between real data and the fake data created by the generator.
\end{itemize}

The training process is a competition. The generator gets better at creating convincing data to fool the Discriminator, while the Discriminator becomes more skilled at telling the real data from the fake. Over time, the generator produces data that is increasingly difficult to distinguish from the real thing.

\subsection{Data Preperation}

Before we start modelling, we must prepare our data for training. While data preparation isn't a one-size-fits-all process, and it varies significantly depending on the dataset at hand, in my case, I discovered that the original dataset, which has 32 columns, introduced unnecessary complexity. While having more data can sometimes be beneficial, too many features can make it harder for a neural network to effectively capture the relationships between variables, increasing both computational load and the risk of overfitting.

To address this, I simplified the PVS dataset by reducing it from 32 columns down to just six key features: \textit{latitude}, \textit{longitude}, \textit{speed}, \textit{acceleration}, \textit{gyro}, and \textit{mag}. I achieved this reduction by taking the absolute mean of the acceleration, gyroscope, and magnetometer sensor readings and combining them, respectively.Table~\ref{table5} presents a sample of the processed dataset.

\begin{table}[h]
\caption{Processed Dataset with Reduced Features}
\label{table5}
\centering
\begin{tabular}{ccccccc}
\hline
\textbf{Lat.} & \textbf{Long.} & \textbf{Speed} & \textbf{Acc.} & \textbf{Gyro} & \textbf{Mag} & \textbf{Road} \\
\hline
-27.7178 & -51.0988 & 0.0091 & 3.4341 & 0.0991 & 24.2361 & 0 \\
-27.7178 & -51.0988 & 0.0091 & 3.4382 & 0.0925 & 24.4036 & 0 \\
-27.7178 & -51.0988 & 0.0091 & 3.4447 & 0.1157 & 24.0074 & 0 \\
-27.7178 & -51.0988 & 0.0091 & 3.4340 & 0.1062 & 24.1384 & 0 \\
-27.7178 & -51.0988 & 0.0091 & 3.4432 & 0.0980 & 24.3430 & 0 \\
\hline
\end{tabular}
\end{table}

In this table, the ``Road'' column represents the type of road surface, encoded as numerical categories: 0 for ``asphalt'', 1 for ``cobblestone'', and 2 for ``dirt'' roads. Encoding the road types transforms categorical data into numerical values that our models can process more efficiently.

\section{Generating Synthetic Data}
Finally, we can start generating data. To create synthetic data for our analysis, I used the Tabular Variational Autoencoder (TVAE) model provided by the Synthetic Data Vault (SDV) library \cite{tvae}. After testing the various models available, I found the TVAE to be most well-suited for our data, configuring it with the following hyperparameters:

\begin{itemize}
    \item \textbf{Epochs:} 200
    \item \textbf{Batch Size:} 500
    \item \textbf{Encoder Layers:} (128, 128)
    \item \textbf{Decoder Layers:} (128, 128)
    \item \textbf{Embedding Dimension:} 128
    \item \textbf{L2 Regularisation Scale:} $1 \times 10^{-5}$
    \item \textbf{Loss Factor:} 2
    \item \textbf{Enforce Min-Max Values:} False
    \item \textbf{Enforce Rounding:} False
\end{itemize}

I trained the TVAE model on a selection of features crucial for road surface classification: \textit{latitude}, \textit{longitude}, \textit{speed}, \textit{acceleration}, \textit{gyro}, \textit{mag} and the target variable \textit{road\_encoded}. These features were chosen for their relevance to the classification task while ensuring that sensitive information, such as exact geographical locations, remained protected to address privacy concerns.

\begin{figure}[h]
    \centering
    \includegraphics[width=0.5\textwidth]{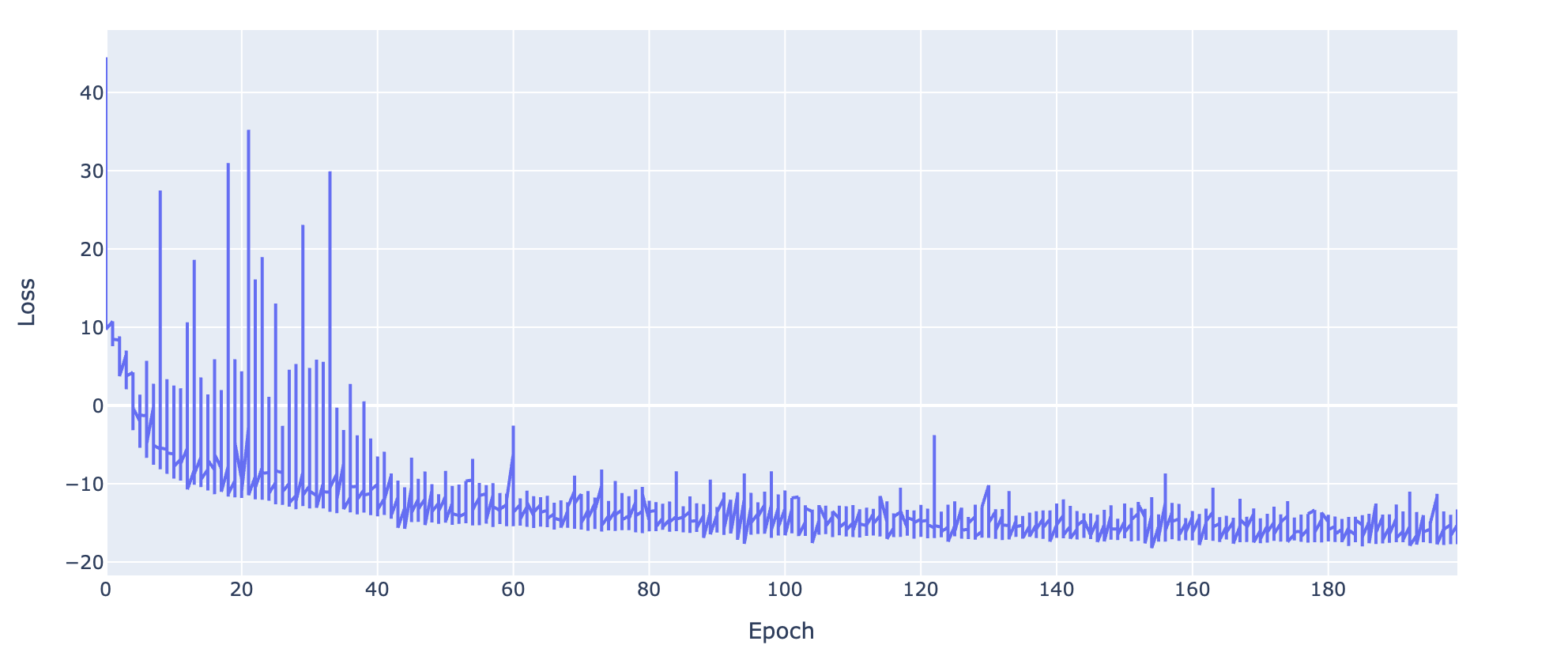}
    \caption{TVAE training loss over 200 epochs.}
    \label{fig:tvae_loss}
\end{figure}

Figure~\ref{fig:tvae_loss} illustrates the training loss over 200 epochs. Initially, the loss fluctuated during the first 50 epochs, which is typical as the model begins to learn the data distribution. By epoch 60, the loss stabilised, indicating that the model had effectively converged. This convergence suggests that the TVAE successfully captured the underlying patterns in the data, balancing reconstruction accuracy with generalisation capabilities.

While the convergence of the training loss is a positive indicator, it is insufficient to assess the overall quality of the synthetic data. To thoroughly evaluate the synthetic data, I generated 144,036 samples—matching the size of the original dataset—and examined three key metrics: \textbf{fidelity}, \textbf{utility}, and \textbf{privacy}.

\subsection{Evaluation: Fidelity}

Fidelity refers to the extent to which the synthetic data replicates the statistical properties of the original dataset. To assess fidelity, I utilized the SDV's evaluation metrics, which provide quantitative scores for the similarity between the real and synthetic data.

The evaluation comprises two main components:

\begin{itemize}
    \item \textbf{Column Shapes Score}: Measures the statistical similarity of individual features (univariate distributions) between the real and synthetic data.
    \item \textbf{Column Pair Trends Score}: Assesses the similarity in relationships between pairs of features (bivariate distributions), effectively capturing the correlations present in the data.
\end{itemize}

The synthetic data achieved the following scores:

\begin{itemize}
    \item \textbf{Column Shapes Score}: 91.12\%
    \item \textbf{Column Pair Trends Score}: 89.07\%
\end{itemize}

An overall average score of 90.10\% was obtained, indicating a high degree of fidelity. These scores suggest that the synthetic data closely mirrors the statistical distributions and relationships found in the original dataset.

To further illustrate the fidelity of the synthetic data, Table~\ref{tab:statistical_comparison} presents a statistical comparison of key features between the real and synthetic datasets.

\begin{table*}[t]
\centering
\caption{Statistical Comparison of Real and Synthetic Data}
\renewcommand{\arraystretch}{1.2} 
\setlength{\tabcolsep}{5pt} 
\small 
\resizebox{\textwidth}{!}{
\begin{tabular}{lcccccc}
\hline
\textbf{Feature} & \textbf{Mean (Real)} & \textbf{Std Dev (Real)} & \textbf{Skewness (Real)} & \textbf{Mean (Synthetic)} & \textbf{Std Dev (Synthetic)} & \textbf{Skewness (Synthetic)} \\ \hline
Latitude        & -27.6946 & 0.0113 & -1.1112  & -27.6942 & 0.0104 & -1.0713 \\
Longitude       & -51.1198 & 0.0111 & 0.7368   & -51.1200 & 0.0095 & 0.5406  \\
Speed           & 9.5569   & 7.7464 & 0.8553   & 11.0330  & 7.6291 & 0.6704  \\
Acceleration    & 4.3620   & 0.9513 & 1.2653   & 4.4054   & 0.7922 & 1.0937  \\
Gyro            & 6.1020   & 3.6026 & 0.6371   & 5.4270   & 2.2154 & -0.0178 \\
Mag             & 25.4643  & 4.0025 & -0.4502  & 25.7967  & 3.2753 & 0.0477  \\ \hline
\end{tabular}
}
\label{tab:statistical_comparison}
\end{table*}

\subsubsection{\textbf{Statistical Comparisons}}

As observed in Table~\ref{tab:statistical_comparison}, each feature's mean and standard deviation in the synthetic data closely align with those of the real data. For instance, the mean latitude in the real data is $-27.6946$, while in the synthetic data, it is $-27.6942$, indicating minimal deviation. Similarly, the standard deviations are comparable across all features, suggesting that the synthetic data captures the variability in the original dataset.

The skewness values, which measure the asymmetry of the distributions, also exhibit close correspondence for most features. Minor differences are present in the \textit{gyro} and \textit{mag} features, where the synthetic data shows slight variations in skewness. These differences, however, are relatively small and do not significantly impact the overall fidelity.

\subsubsection{\textbf{Correlation Analysis}}

\begin{figure}[h]
    \centering
    \includegraphics[width=0.5\textwidth]{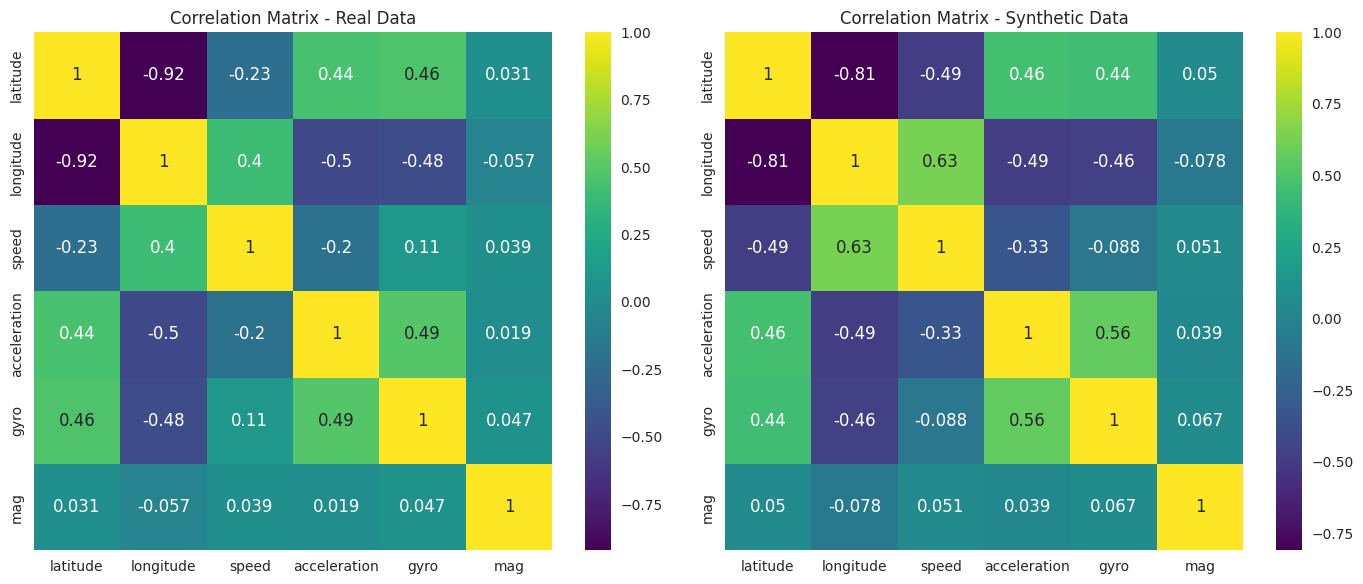}
    \caption{Correlation matrix of the real data.}
    \label{fig:real_synthetic_correlation}
\end{figure}

To assess how well the synthetic data preserves the relationships between features, I examined the correlation matrices of both datasets. Figure ~\ref{fig:real_synthetic_correlation} displays the correlation matrices for the real and synthetic data.

The correlation matrices reveal that the synthetic data effectively replicates the inter-feature relationships present in the real data. Notably, the strong negative correlation between \textit{latitude} and \textit{longitude} is preserved, with coefficients of $-0.92$ in the real data and $-0.81$ in the synthetic data. Additionally, the correlations between \textit{acceleration} and other features, such as \textit{gyro} and \textit{speed}, are consistently maintained.

\subsubsection{\textbf{Distribution Comparison Using KDE Plots}}

\begin{figure*}[htbp] 
    \centering
    \begin{subfigure}[b]{0.3\textwidth}
        \includegraphics[width=\textwidth]{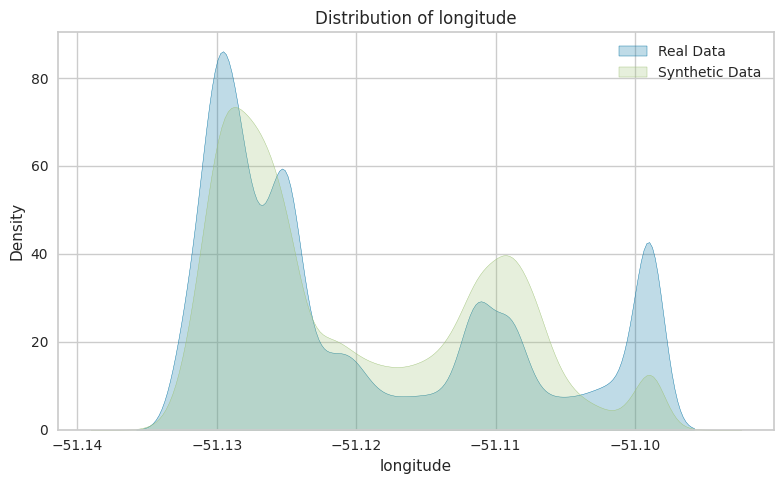}
        \caption{Longitude}
        \label{fig:kde_longitude}
    \end{subfigure}
    \hfill
    \begin{subfigure}[b]{0.3\textwidth}
        \includegraphics[width=\textwidth]{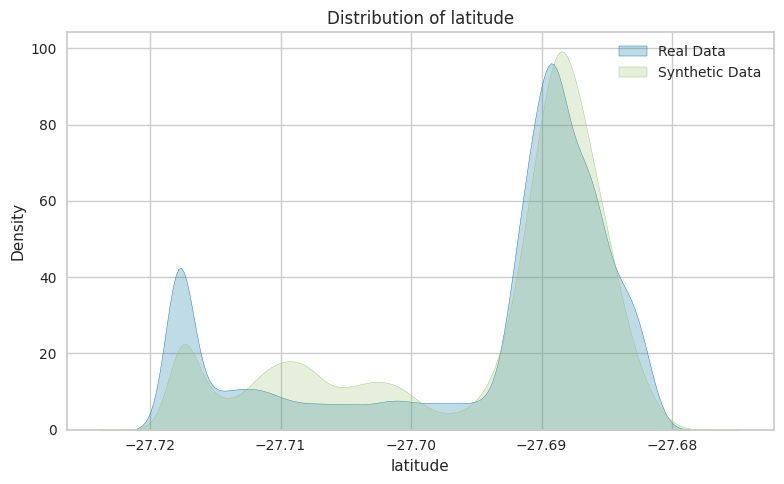}
        \caption{Latitude}
        \label{fig:kde_latitude}
    \end{subfigure}
    \hfill
    \begin{subfigure}[b]{0.3\textwidth}
        \includegraphics[width=\textwidth]{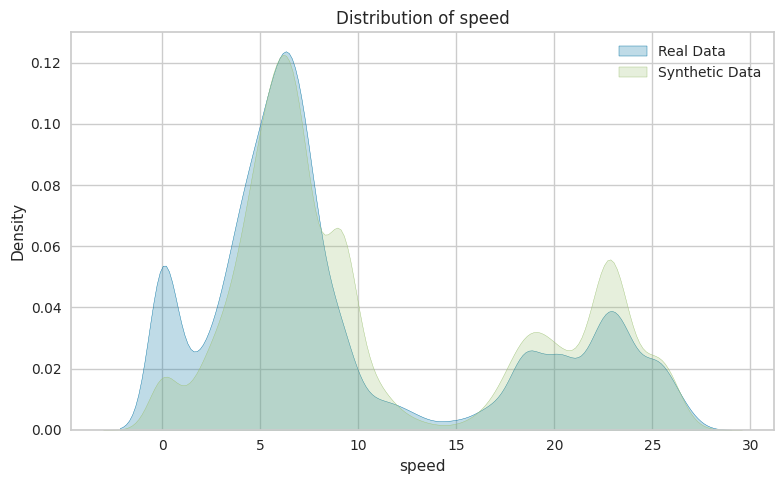}
        \caption{Speed}
        \label{fig:kde_speed}
    \end{subfigure}
    
    \vspace{1em} 

    \begin{subfigure}[b]{0.3\textwidth}
        \includegraphics[width=\textwidth]{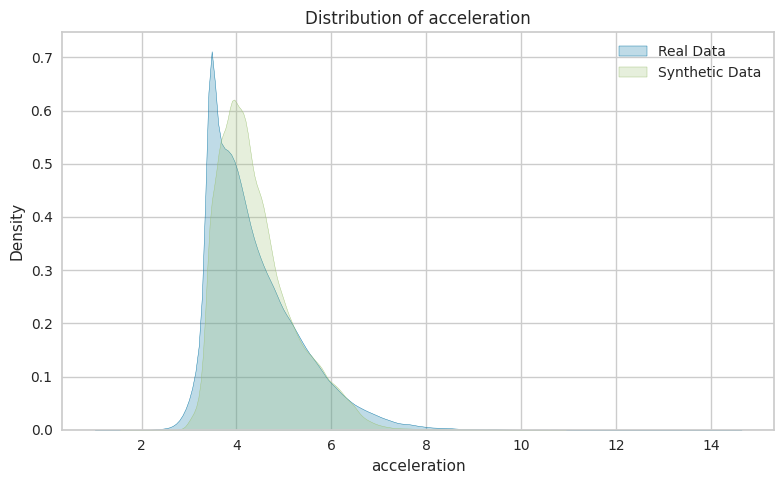}
        \caption{Acceleration}
        \label{fig:kde_acceleration}
    \end{subfigure}
    \hfill
    \begin{subfigure}[b]{0.3\textwidth}
        \includegraphics[width=\textwidth]{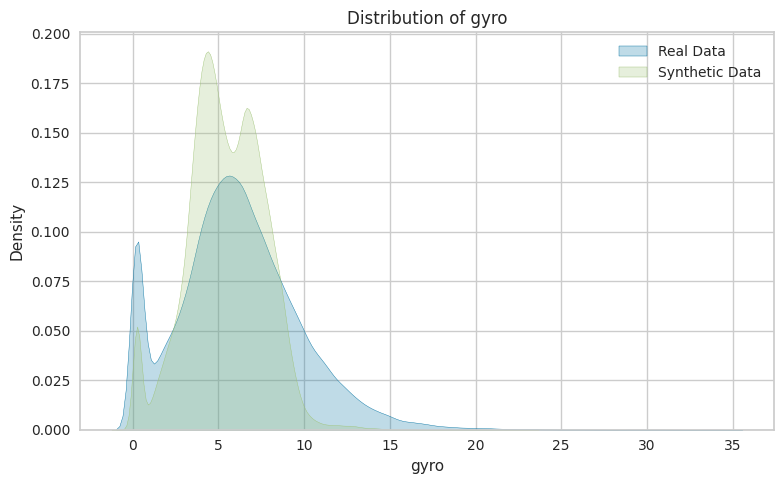}
        \caption{Gyro}
        \label{fig:kde_gyro}
    \end{subfigure}
    \hfill
    \begin{subfigure}[b]{0.3\textwidth}
        \includegraphics[width=\textwidth]{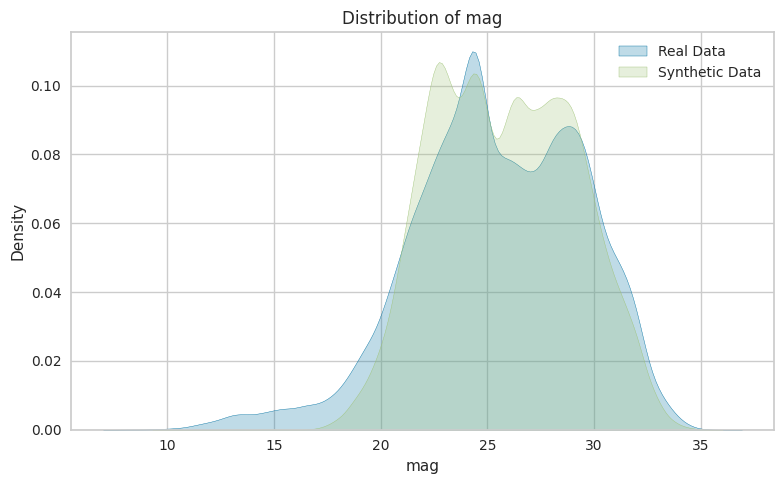}
        \caption{Magnetic Field Magnitude}
        \label{fig:kde_mag}
    \end{subfigure}
    
    \caption{KDE plots comparing the distributions of features in the real and synthetic data. The blue curve represents the real data, while the orange curve represents the synthetic data.}
    \label{fig:kde_plots}
\end{figure*}

To visualize the distributions of individual features, I generated Kernel Density Estimation (KDE) plots for both the real and synthetic data. Figure~\ref{fig:kde_plots} showcases the KDE plots for selected features.

The KDE plots demonstrate that the synthetic data closely follows the distributions of the real data across all features. For example, the distribution of \textit{speed} in the synthetic data aligns well with that of the real data, capturing both the central tendency and the spread. We observe slight deviations in the tails of the distributions for certain features, such as \textit{gyro}, but these differences are minor.

High fidelity scores and the statistical and visual analyses indicate that the synthetic data generated by the TVAE model is a faithful representation of the real dataset. This suggests that the model effectively learned the underlying data distribution and can produce synthetic data that maintains the essential characteristics of the original data.

\subsection{Evaluation: Utility}
Utility refers to how effectively the synthetic data can perform the intended tasks of the original dataset. Here, the primary objective is to classify road surfaces into three categories: \textit{asphalt}, \textit{cobblestone}, and \textit{dirt}, using features such as \textit{latitude}, \textit{longitude}, \textit{speed}, \textit{acceleration}, \textit{gyro}, and \textit{magnetic field magnitude}.

Classification models were trained on both real and synthetic datasets to assess utility, and their performance was compared using the original real data. Evaluation metrics included accuracy, precision, recall, and F1-score to provide a comprehensive understanding of each model's effectiveness.

\subsubsection{\textbf{Model Trained and Tested on Real Data}}

Using the PyCaret library, multiple classification models were trained and evaluated on the real dataset. The best-performing model was the K-Nearest Neighbors (KNN) classifier, achieving an accuracy of 96.89\%. Table~\ref{tab:real_data_models} summarizes the performance metrics of the top models trained on the real data.

\begin{table}[htbp]
\centering
\caption{Performance of Classification Models Trained on Real Data}
\renewcommand{\arraystretch}{1.1} 
\setlength{\tabcolsep}{3pt} 
\scriptsize 
\begin{tabular}{llccccc}
\hline
\textbf{Model} & \textbf{Description} & \textbf{Accuracy} & \textbf{AUC} & \textbf{Precision} & \textbf{Recall} & \textbf{F1-score} \\
\hline
\textbf{KNN} & K-Nearest Neighbors & 0.9689 & 0.9970 & 0.9689 & 0.9689 & 0.9689 \\
\textbf{DT} & Decision Tree & 0.9677 & 0.9989 & 0.9676 & 0.9677 & 0.9675 \\
\textbf{ET} & Extra Trees & 0.9677 & 0.9989 & 0.9676 & 0.9677 & 0.9675 \\
\textbf{RF} & Random Forest & 0.9676 & 0.9989 & 0.9676 & 0.9676 & 0.9675 \\
\hline
\end{tabular}
\label{tab:real_data_models}
\end{table}

The KNN classifier's high accuracy demonstrates the robustness of real data in distinguishing different road surfaces. Overall, the models exhibit strong performance across all evaluation metrics, highlighting the effectiveness of the selected features for classification.

\subsubsection{\textbf{Model Trained on Synthetic Data and Tested on Real Data}}

Using the synthetic data generated by the TVAE model, classification models were trained and tested on the original real data to evaluate generalizability to real-world scenarios.

\begin{figure}[h]
    \centering
    \includegraphics[width=0.5\textwidth]{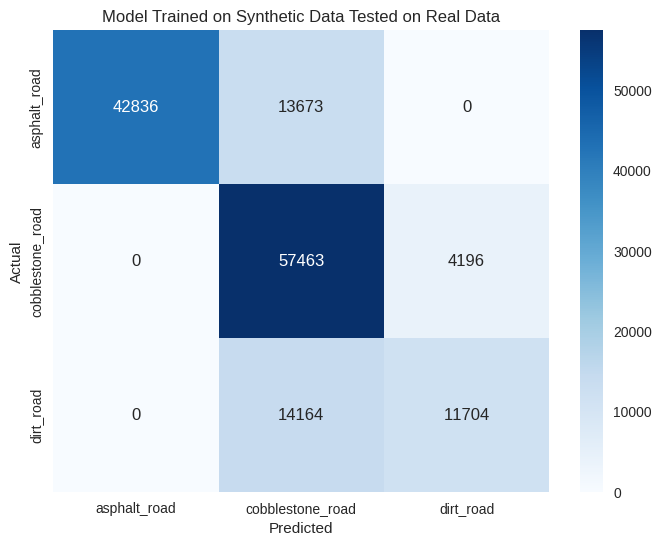}
    \caption{Confusion Matrix: XGBoost Model Trained on Synthetic Data Tested on Real Data}
    \label{fig:confusion_matrix_synthetic}
\end{figure}

The best-performing model in this evaluation was the Extreme Gradient Boosting (XGBoost) classifier, achieving an accuracy of 77.76\% when tested on real data. Table \ref{tab:classification_report} and Figure \ref{fig:confusion_matrix_synthetic} present the detailed classification report and confusion matrix, respectively.

\begin{table}[h!]
\centering
\caption{Classification Report}
\begin{tabular}{|l|c|c|c|c|}
\hline
              & \textbf{Precision} & \textbf{Recall} & \textbf{F1-score} & \textbf{Support} \\ \hline
\textbf{Class 0} & 1.00             & 0.76           & 0.86              & 56,509           \\ \hline
\textbf{Class 1} & 0.67             & 0.93           & 0.78              & 61,659           \\ \hline
\textbf{Class 2} & 0.74             & 0.45           & 0.56              & 25,868           \\ \hline
\textbf{Accuracy} & \multicolumn{4}{c|}{0.78 (on 144,036 samples)} \\ \hline
\textbf{Macro avg} & 0.80            & 0.71           & 0.73              & 144,036          \\ \hline
\textbf{Weighted avg} & 0.81         & 0.78           & 0.77              & 144,036          \\ \hline
\end{tabular}
\label{tab:classification_report}
\end{table}

\subsubsection{\textbf{Comparison of Models}}

\begin{figure}[h]
    \centering
    \includegraphics[width=0.5\textwidth]{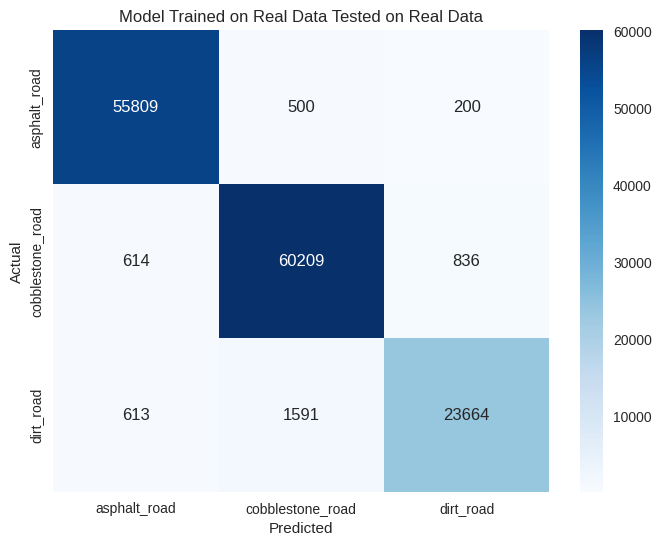}
    \caption{Confusion Matrix: KNN Model Trained on Real Data Tested on Real Data}
    \label{fig:confusion_matrix_real}
\end{figure}

As shown in Figure \ref{fig:confusion_matrix_real}, the KNN model demonstrates high accuracy across all classes, with minimal misclassifications. The model effectively identified road types with high precision and recall. In contrast, the XGBoost model trained on synthetic data often misclassified \textit{asphalt} and \textit{dirt} instances as \textit{cobblestone}, suggesting that the synthetic data did not fully capture the distinct features necessary for accurately differentiating road types. 

Moreover, class imbalance in the dataset, particularly fewer instances of \textit{dirt}, adversely affected model performance for the minority class. The models tended to bias towards the majority classes, leading to lower recall for \textit{dirt} road types.

Overall, the utility evaluation reveals that while models trained on synthetic data can achieve moderate performance, they do not match the accuracy of those trained on real data. Although the synthetic data preserves statistical properties, it may not fully encapsulate the complex relationships required for precise classification in this application.

These findings suggest that while synthetic data generated by the TVAE model holds value—particularly for preliminary analyses or privacy-sensitive scenarios—it is currently limited in training high-performance predictive models.

\subsection{Evaluation: Privacy}
Our final evaluation metric is privacy. Privacy is critical when generating and utilising synthetic data, especially when dealing with sensitive information such as geographical coordinates that could potentially identify individuals. In our context, privacy refers to the ability of the synthetic data to protect the original vulnerable data—specifically, the \textit{longitude} and \textit{latitude} features—from being used to identify the driver or reveal precise travel routes.

While synthetic data is often assumed to enhance privacy, as mentioned at the start of this paper, it is not inherently private. Therefore, evaluating how well the synthetic data masks the original data points is crucial for determining its viability in applications where privacy is a concern.

\subsubsection{\textbf{Assessing Privacy}}

The challenge with privacy is that it is hard to quantify. How does one evaluate how vulnerable something is or isn't to revealing sensitive information? Therefore, the only way we can ensure our data is private is by evaluating how well we can reconstruct sensitive attributes from our given dataset.

Given the geographical nature of our data, we can take a practical approach to assessing privacy by aiming to re-identify the driver's original path from the synthetic data.

\begin{figure*}[htbp]
    \centering
    \begin{subfigure}[b]{0.45\textwidth}
        \includegraphics[width=\textwidth]{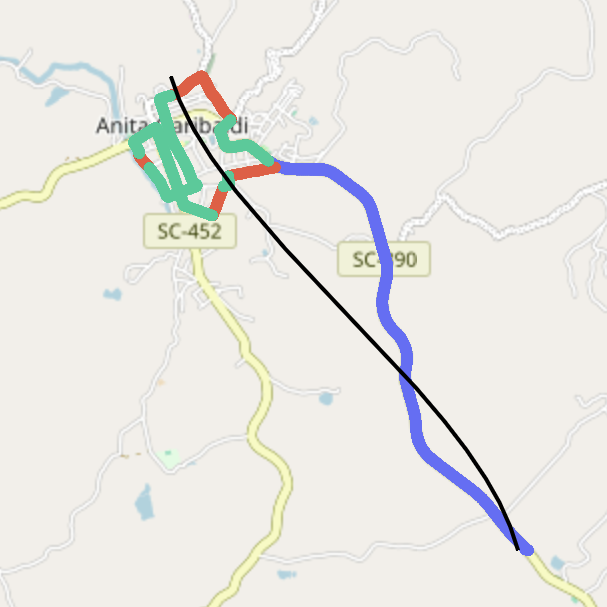}
        \caption{Original Data (144036 rows)}
        \label{fig:original_geo_plot}
    \end{subfigure}
    \hfill
    \begin{subfigure}[b]{0.45\textwidth}
        \includegraphics[width=\textwidth]{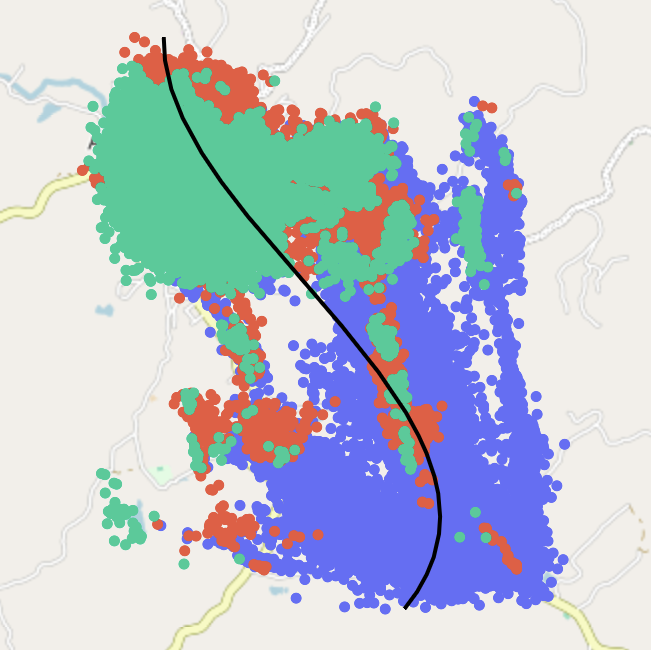}
        \caption{Synthetic Data (144036 rows)}
        \label{fig:synthetic_geo_plot}
    \end{subfigure}
    \caption{Geographical data comparison between original and synthetic datasets. Points are colour-coded by road type: red (dirt), green (cobblestone), blue (asphalt).}
    \label{fig:geo_comparison}
\end{figure*}

By visually comparing plots of the \textit{longitude} and \textit{latitude} data in both the real and synthetic datasets, Figure~\ref{fig:geo_comparison} presents a side-by-side comparison of the geographical data from the original and synthetic datasets. In these plots, data points are colour-coded based on the road surface type (\textit{asphalt road}, \textit{cobblestone road}, \textit{dirt road}).

From Figure~\ref{fig:geo_comparison}, several specific observations can be made:

\begin{itemize}
    \item \textbf{General Route Patterns Preserved}: The synthetic data maintains the overall route patterns observed in the original dataset. For instance, it still reflects that the driver travelled along the SC-390 road towards Anita Garibaldi. The general trajectory and orientation of the points align with major roads.

    \item \textbf{Specific Locations Masked}: In the original data, we can pinpoint locations within Anita Garibaldi where the driver stopped or passed through. The synthetic data presents only clusters of points around the town, concealing precise paths and stopping points. This prevents the identification of specific addresses or landmarks that the driver may have stopped at, thereby enhancing privacy by reducing the risk of re-identifying the driver's exact route.

    \item \textbf{Starting and Ending Points Obscured}: The original dataset reveals where the driver started and ended their journey, which could be sensitive information like a home or workplace. The synthetic data generalises these points, making it unclear where the trip began or concluded. This obscurity protects the driver's privacy by preventing the tracing of trips back to specific locations.

    \item \textbf{Linear Trends Maintained}: By plotting linear regression lines on both datasets, we observe that the synthetic data maintains similar linear trends to the real data. This indicates that while individual points are altered, the underlying relationships between geographical features are preserved.
    
\end{itemize}

These observations allow us to conclude that the synthetic data effectively maintains the privacy of the data it has been trained on. Comparing the similarity of the regression lines shows us that the geographical features of the dataset have been preserved, indicating maintained fidelity.

It is worth noting that this increased privacy has subsequently resulted in a fall in the utility of the data for road type classification as visually from \ref{fig:geo_comparison} and from our results in \ref{tab:classification_report} it is clear to see the existence of poorly distributed road types relative to their original positioning.

Our findings, therefore, suggest that while synthetic data generated by the TVAE model provides sufficient privacy and statistical similarity for the geographical data, the increased noise within the data hindered its effectiveness in classifying road types.

\section{Conclusion: What Did We Learn?}

In this paper, we explored the role of synthetic data in enhancing privacy for smart vehicles. Beginning with an introduction to synthetic data and its historical development, we highlighted its potential benefits and common misconceptions. Recognising the increasing risks of information leakage from smart vehicles, we addressed the specific problem.

To systematically assess these risks, I proposed a comprehensive taxonomy of in-vehicle sensors \ref{tab:taxonomy} based on their vulnerability to information leakage. This taxonomy categorised 14 key signals into high, medium, and low priority, identifying GPS data and camera feeds as the most susceptible to privacy breaches. By understanding which sensors posed the greatest risks, we focused on protecting the most sensitive data.

Focusing on GPS data due to its high vulnerability, we then needed to decide which means of synthetic data generation we would follow. Having created a high-level flow chart depicting the various methods, models, and their use cases, we settled on the Tabular Variational Autoencoder (TVAE) model using the Passive Vehicular Sensor dataset. The goal was to generate synthetic data that preserved the original dataset's statistical properties and road type classification accuracy while mitigating the potential to profile the driver's driving pattern.

We trained our generator and evaluated the synthetic data based on three critical criteria: fidelity, utility, and privacy. Our findings indicated that the synthetic data achieved high fidelity, effectively replicating the statistical distributions and correlations of the original data by around 90\%. However, when assessing utility, models trained on the synthetic data exhibited a 20\% reduction in accuracy when it came to classifying road surface types compared to those trained on real data. This reinforced the known trade-off that exists between privacy and utility.

Importantly, the synthetic data enhanced privacy by obscuring exact geographical coordinates, thereby reducing the risk of re-identification and unauthorised tracking. This outcome underscores the potential of synthetic data to protect individual privacy in smart vehicles without entirely sacrificing data utility.

Our results, therefore, emphasise the inherent challenges in balancing privacy and utility when generating synthetic data. 
By leveraging our taxonomy to identify and prioritise vulnerabilities, future work could explore hybrid approaches, such as incorporating a mix of real and synthetic data points, to assist in enabling the continued advancement of smart vehicle technologies while safeguarding individual privacy.

\bibliographystyle{IEEEtran}
\bibliography{references}

\end{document}